\documentstyle[12pt]{article}
\begin{document}
%
\def\lsim{\:\raisebox{-0.5ex}{$\stackrel{\textstyle<}{\sim}$}\:}
\def\gsim{\:\raisebox{-0.5ex}{$\stackrel{\textstyle>}{\sim}$}\:}
\def\bib{\bibitem}
\def\d{\dagger}
\def\hw {\hbar \omega}
\def\ad{ {\bf a}^{\dagger}}
\def\ba#1{\begin{array}{#1}}
\def\ea{\end{array}}
\def\bin#1#2{\frac{#1!}{(#1-#2)! #2!}}
\def\hre#1#2{\frac{#1!}{(#1-2#2)! #2!}}
\def\hreg#1#2#3{\frac{#1!}{#2! (#1-#3 #2)!}}
\def\nsp{\negthinspace}
\def\b#1#2{\left(\nsp \begin{array}{r}#1\\#2\end{array} \nsp \right)}
\def\hr#1#2{\left[\nsp \begin{array}{c} #1\\ #2 \end{array} \nsp \right]}
\def\hrg#1#2#3{\left[\nsp\ba{c}#1\\#2\ea \nsp \right]_{#3}}
\def\above#1#2{{\stackrel {{\textstyle #1}}{\scriptstyle { #2}}}}
\def\pb{\beta^{'} }
\def\be{\begin{equation}}
\def\ee{\end{equation}}
\def\br{\begin{eqnarray}}
\def\er{\end{eqnarray}}
\def\x{\times}
\def\hs{{\cal H}}
\def\go{\rightarrow  }
\def\abs#1{{\mid #1 \mid }}
\def\la{\langle }
\def\kg{\preceq }
\def\ra{\rangle }
\def\lb#1{{\label{eq:#1} }}
\def\rf#1{{(\ref{eq:#1}) }}
\def\dox{$\times$ }
\def\dodt{{$\bullet $ }}
\def\bra#1{\langle #1|}
\def\ket#1{|#1 \rangle}
\def\bra#1{\langle #1|}
\def\ex#1#2{\langle #1 | #2 | #1 \rangle }
\def\sp#1{\langle #1 \rangle }
\def\ov#1#2{\langle #1 | #2  \rangle }
\def\exp#1#2#3{\langle #1 | #2 | #3 \rangle }
\def\L {{\Lambda}}
\def\T {{\Theta}}
\def\U {{\Upsilon}}
\def\G {{\Gamma}}
\def\O {{{\cal O}}}
\def\M {{{\cal M}}}
\def\N {{{\cal N}}}
\def\Q {{{\cal Q}}}
\def\S {{{\cal S}}}
\def\up{u_{p}}
\def\vp{v_{p}}
\def\un{u_{n}}
\def\vn{v_{n}}
\def\rop{\rho_{p}}
\def\ron{\rho_{n}}
\def\sip{\sigma_{p}}
\def\sin{\sigma_{n}}
\def\upp{u_{p'}}
\def\vpp{v_{p'}}
\def\unp{u_{n'}}
\def\vnp{v_{n'}}
\def\ropp{\rho_{p'}}
\def\ronp{\rho_{n'}}
\def\Bra#1{\langle #1||}
\def\Ket#1{||#1 \rangle}
\def\ubp{\bar{u}_{{\rm p}}}
\def\vbp{\bar{v}_{{\rm p}}}
\def\ubn{\bar{u}_{{\rm n}}}
\def\vbn{\bar{v}_{{\rm n}}}
\def\up{u_{{\rm p}}}
\def\vp{v_{{\rm p}}}
\def\un{u_{{\rm n}}}
\def\vn{v_{{\rm n}}}
\def\rop{\rho_{{\rm p}}}
\def\ron{\rho_{{\rm n}}}
\def\ubpp{\bar{u}_{{\rm p}'}}
\def\vbpp{\bar{v}_{{\rm p}'}}
\def\ubnp{\bar{u}_{{\rm n}'}}
\def\vbnp{\bar{v}_{{\rm n}'}}
\def\upp{u_{{\rm p}'}}
\def\vpp{v_{{\rm p}'}}
\def\unp{u_{{\rm n}'}}
\def\vnp{v_{{\rm n}'}}
\def\ropp{\rho_{{\rm p}'}}
\def\ronp{\rho_{{\rm n}'}}
\def\beg#1#2{\left(\nsp \begin{array}{r}#1\\#2\end{array} \nsp \right)}
\begin{titlepage}
\pagestyle{empty}
\baselineskip=21pt
\vskip .2in
\begin{center}
{\large{\bf Double Beta Decay in pn-QRPA Model with Isospin and SU(4)
Symmetry Constraints }}
\end{center}
\vskip .1in
\begin{center}
F. Krmpoti\'{c}$^{\d}$

{\small\it Departamento de F\'\i sica, Facultad de Ciencias Exactas,\\
 Universidad Nacional de La Plata, C. C. 67, 1900 La Plata, Argentina}\\

S. Shelly Sharma$^*$

{\small \it  Departamento de F\'\i sica, Universidade Estadual de Londrina,
Londrina, Parana,  and\\
 Instituto de F\'\i sica Te\'orica, Rua Pamplona-145
 CEP 01405 - S\~ao Paulo, Brazil}

%
\end{center}
\centerline{ {\bf Abstract} }
\baselineskip=18pt
The transition matrix elements for the $0^{+}\rightarrow 0^{+}$ double beta
decays  are calculated for $^{48}Ca$, $^{76}Ge $, $^{82}Se$, $^{100}Mo$,
$^{128}Te$ and $^{130}Te$ nuclei, using a ${\delta}$-interaction.  As a
guide, to fix the particle-particle interaction strengths, we exploit the fact
that the missing symmetries of the mean field approximation are restored in
the random phase approximation by the residual interaction.
Thus, the $T=1$, $S=0$ and $T=0$, $S=1$ coupling strengths have been estimated
by invoking the partial restoration of the isospin and Wigner SU(4) symmetries,
respectively.
When this recipe is strictly applied, the calculation is consistent
with the experimental limit for the $2\nu$ lifetime of $^{48}Ca$ and  it
also correctly reproduces the $2\nu$ lifetime of $^{82}Se$.
In this way, however, the two-neutrino matrix elements for the remaining nuclei
are either underestimated (for $^{76}Ge$ and $^{100}Mo$) or overestimated (for
$^{128}Te$ and $^{130}Te$) approximately by a factor of 3.
With a comparatively small variation ($<10\%$) of the spin-triplet
parameter, near the value suggested by the SU(4) symmetry, it
is possible to reproduce the measured $T_{1/2}^{2\nu}$ in all the cases.  The
upper limit for the effective neutrino mass, as obtained from the theoretical
estimates of $0\nu$ matrix elements, is $<m_{\nu}>\cong 1$ eV.  The
dependence of the nuclear matrix elements on the size of the configuration
space has been also analyzed.

\vspace{0.4in}
\noindent
$^{\d}$Fellow of the CONICET from Argentina\\
$^*$Financial support from CNPq, Brazil is acknowledged
\end{titlepage}
\baselineskip=18pt
\newpage
\section {Introduction}
\label{sec1}
The  odd-odd  isobar,  within  the isobaric triplet $(N,Z)$, $(N-1,Z+1)$,
$(N-2,Z+2)$, generally has a higher mass than its even-even neighbors due to
the pairing energy. As such, two consecutive $\beta$ decays
are energetically forbidden.  Yet, the initial
nucleus $(N,Z)$ may decay  to the  final nucleus  $(N-2,Z+2)$ through virtual
excitations of the intermediate  nucleus  $(N-1,Z+1)$, causing a double
beta ($\beta \beta$) decay process. In case the lepton number  is strictly
conserved the neutrino is a Dirac fermion ($\nu \neq \tilde{\nu}$) and
the two-neutrino mode ($\beta \beta_{2\nu}$) is the only  possible mode
of  disintegration. On the other hand, if this conservation
is violated, the neutrino is a Majorana  particle ($\nu = \tilde{\nu}$)
and the neutrinoless double beta decay  ($\beta \beta_{0\nu}$) also can occur.

Both the $\beta \beta_{2\nu}$ and $\beta \beta_{0\nu}$-decay processes have
attracted   much attention during the last decade. This is  mainly because
the  neutrinoless decay mode constitutes the most critical touchstone for
various gauge models that go  beyond  the standard $SU(2)_{L}\times U(1)$
gauge model of the electroweak interaction.
The $\beta \beta_{0\nu}$ decay rate depends on several  unknown parameters
(neutrino mass, majoron coupling, the coupling constants of the right-handed
components of the weak Hamiltonian,  etc.)  and the only way to put  these
in  evidence is by having  sufficient command over the nuclear structure.
It is precisely at  this point that the $\beta \beta_{2\nu}$ decay mode is
important. A comparison between experiment and theory for the
$\beta \beta_{2\nu}$-decay, provides a measure of confidence one may have
in the nuclear wave functions employed for extracting the unknown parameters
from $\beta \beta_{0\nu}$ lifetime measurements.

In recent years, following the work of Vogel and Zirnbauer \cite{Vog86},
the proton-neutron (pn)  quasiparticle  random phase approximation (QRPA)
has turned out to be the most popular model for calculating the nuclear wave
functions involved in the $\beta \beta$-transitions
\cite{Civ87,Tom87,Eng88,Eng88a,Mut88,Mut88a,Mut89,Hir90,Hir90a,Hir90b,Krm90,Pan90,Suh90,Sta90,Sta90a,Civ91,Mut91,Suh91,Suh91a,Tom91,Krm92}.
The QRPA calculations have shown that the ground state correlations (GSC),
induced by the residual pn interaction in the particle-particle (PP)
channel, play an essential role in quenching the $\beta \beta_{2\nu}$ decay
probabilities. Regarding the  sensitivity   of   $\beta \beta_{0\nu}$
nuclear moments, ${\cal M}_{0\nu}$, to  the  PP interaction, no consensus
has been reached so far. The results obtained by the Pasadena group
\cite{Eng88}, with a residual $\delta$ interaction, seem to show  that
${\cal M}_{0\nu}$  are rather sensitive to  this  force and  the $^{76}Ge$
lifetime measurement \cite{Bal92} $(T_{1/2}^{0\nu}>1.4\times 10^{24} yr)$
yields the upper bound on the effective neutrino mass, $<m_{\nu}>$,
varying between 4.4 and 10 eV
(see table \ref{table1}). On the other hand, the calculations done by
T\"{u}bingen and Heidelberg groups with a realistic interaction (the
G-matrix of the Paris and Bonn potentials)
\cite{Tom87,Mut89,Mut91,Suh91a}, suggest that the ${\cal M}_{0\nu}$ moments
are to some
extent insensitive to the  PP channel, which results in a smaller neutrino
mass with
more stringent upper limit (between 2.3 eV and 3.1 eV) from the
same datum \cite{Bal92}. The difference between the three sets of calculations
lies in the  importance  attached  to  the role of destructive interference
between the allowed $(L=0)$ and the forbidden $(L>0)$ virtual transitions.
There is a general agreement that, for the physically acceptable values of PP
coupling strength, $L=0$ contributions are relatively small and those  from
$L>0$ processes are significant. But, whereas the Pasadena group believes
that the allowed matrix elements are strongly canceled by the forbidden ones,
the other two groups claim the cancellation effect to be of minor importance.
The exceptions are the results for the ${\cal M}_{0\nu}$ moments in $^{48}Ca$,
$^{100}Mo$ and $^{128}Te$: for the first two nuclei very small nuclear matrix
elements have been obtained by Heidelberg group, suggesting that the $L=0$
matrix elements are sizable, while for the last nucleus all three groups
obtain similar results.

An important question in the QRPA calculations is, how to fix the PP
interaction strengths within the $S=0$, $T=1$ and $S=1$, $T=0$ channels?
Several attempts have been made to calibrate the last one using the
experimental data for  individual Gamow-Teller (GT) positron decays
\cite{Eng88,Mut88,Mut88a}. The weak point of this procedure is that
the distribution of the $\beta^{+}$ strength among low-lying states in
odd-odd nuclei is certainly affected by the charge-conserving vibrations,
which are not included in the QRPA.
\footnote{The single beta transitions $^{100}Tc\rightarrow{^{100}Mo}$ and
$^{100}Tc \rightarrow{^{100}Ru}$ have been discussed recently in the standard
QRPA \cite{Gri92}. However, from the analysis of the structure of the triplet
of low-lying states in $^{101}Mo$ (cf. ref. \cite{Sey91}), it can be inferred
that the collective degrees of freedom should play a very important role in
these decays.}
In the present work we
resort to the restoration of the isospin and spin-isospin SU(4) symmetries
to gauge, respectively, the $S=0$, $T=1$ and $S=1$, $T=0$ strengths in the PP
channel \cite{Hir90}.  Unlike the method mentioned above, this procedure
involves the  total Fermi (F) and GT strengths, which dependent
of  the charge-conserving vibrations only very weakly.  We are aware, however,
that the SU(4) symmetry is badly broken in heavier nuclei like those
considered here.  As such, before proceeding further, it is necessary
to specify what we mean by reconstruction of isospin and SU(4)
symmetries in the context of our calculation.

For a system with $N\neq Z$, the isospin and spin-isospin symmetries are
violated in  the mean field approximation, even if the nuclear
hamiltonian commutes with the corresponding excitation operators
$\beta^{\pm}$ (${\rm t}_{\mp}$ and $\sigma {\rm t}_{\mp}$).
But, Thouless \cite{Tho60} and Brown \cite{Bro64} have shown that when
a non-dynamical violation occurs in the Hartree-Fock (HF) solution, the RPA
induced GSC can be invoked to restore the symmetry \cite{Eng70,Lee71,Lan80}.
There are subtleties involved in the restoration mechanism: the GSC are not
put in evidence explicitly, but only implicitly via their effects on the
one-body moments $\beta^{\pm}$ between the ground state and the excited
states. Besides, for the F excitations and when the isospin non-conserving
forces are absent, a self-consistent inclusion of the GSC leads to the
following:\\
i) all the $\beta^{-}$  strength is concentrated in the isobaric analog state
(IAS), and\\
ii) the $\beta^{+}$ spectrum, which in RPA can be viewed as an extension of
the $\beta^{-}$ spectrum to negative energies, is totally quenched.\\
The extent to which the above conditions are fulfilled may be taken as a
measure of the isospin symmetry restoration.
\footnote{We may point out that it is not
possible to obtain similar results in the Tamm-Dancoff approximation (TDA)
where the GSC are neglected. As a matter of fact within the TDA the
$\beta^{-}$  strength is always fragmented and the perturbed $\beta^{+}$
strength remains equal to its unperturbed value.}
  Lee \cite{Lee71} has shown that even when the isospin
symmetry is  dynamically broken by the Coulomb force, the RPA induced
GSC result in a near restoration of the isospin symmetry.
Besides being spontaneously broken by the HF approximation,
the SU(4) symmetry is also dynamically torn down by the spin-orbit field and
the supermultiplet-destroying residual interactions.
But, as argued below, the last two effects have a tendency to cancel each
other.

In the mean field approximation (and due to the spin-orbit splitting) the
$\sigma {\rm t}_{-}$ operator has roughly equal strengths for the spin flip as
compared to non-spin flip transitions.
Charge exchange reactions \cite{Bai80,Hor80,Hor81} have revealed, still, that
the spin-isospin residual forces can to some extent overcome
this SU(4) symmetry breaking  by transferring a substantial
amount of the non-spin flip strength to higher energies (and build up in
this way the GT resonance).  In fact, the observed  mean
energy differences between the GT and F resonances can be accounted for by
the relation \cite{Nak82}
\[
E_{GT}-E_F=\left(26 A^{-1/3}-18.5\frac{N-Z}{A}\right) MeV,
\]
and the displacement of the GT resonance towards the IAS
with increasing $N-Z$ may be interpreted as the effect of the residual
interaction \cite{Nak82,Gap81,Suz82}.
As a matter of fact, the first term has the mass dependence of the mean
spin-orbit splitting, $\Delta_{{\it l}s}$, and within  a schematic TDA for
the $\delta$ interaction (that we use in the present work) one gets
\cite{Nak82}
\[
E_{GT}-E_F=\left[\Delta_{{\it l}s}-\left(v_t-v_s\right)
\frac{N-Z}{2A}\right]\,MeV,
\]
where $v_s$ and $v_t$ are, respectively, the singlet and the triplet
coupling constants. It should be noticed that the GSC are likely to alter
this result very little. But, within the RPA the $\sigma {\rm t}_{+}$
transition
strength  is strongly quenched and the GT resonance is somewhat narrowed,
as compared with the TDA results \cite{Krm90}.
As such the global effect of the pn residual interaction on the GT strength
is qualitatively similar to the corresponding effect on the F strength,
and we say that the SU(4) symmetry is partially  restored.
Finally a word  of caution, the closeness of GT resonance to IAS does not
guarantee a complete reconstruction of the SU(4) symmetry.
\footnote {The problem of reconstruction of the Wigner SU(4)  symmetry
and its relationship with the proton-neutron interaction, has been discussed
earlier in the context of the  GT transitions by Fujita and Ikeda \cite{Fuj65}
and by Bohr and Mottelson \cite{Boh69}. Recently Bernabeu et
al. \cite{Ber90} have shown that treating the SU(4) symmetry-breaking part of
the interaction in perturbation theory leads to $\M_{2\nu}$ moments of
correct order of magnitude.  Several other manifestations of the SU(4)
symmetry in nuclei have also been reported
\cite{Gap82,Vog88,Hin88,Mai88,Zhe89,Mut92,Vog93}.}
For example,  the observed GT strength in $^{208}Pb$ is in a single continuous
peak located at the energy of the IAS, but simultaneously the GT resonance is
rather broad ($\approx 4 MeV$; see ref. \cite{Hor80}).

In the present work we perform a detailed analysis of the
$\beta \beta_{0\nu}$ decay rates for several nuclei, always focussing our
attention on the $<m_{\nu}>$ term (mass mechanism). The term containing the
right-handed leptonic currents (RHC mechanism) is not
considered. Besides that, no reference is made to the
neutrinoless $\beta \beta$ decays with single
\cite{Gel81,Geo81}, $\beta \beta_{0\nu,B}$, and double \cite{Moh88},
$\beta \beta_{0\nu,2B}$,  majoron emissions. The $\beta \beta_{0\nu,B}$
decay has been ruled out by the measurement of the Z width \cite{Abr89},
as shown by Gonzalez-Garcia and Nir \cite{Gon89}, whereas no data is available
for the $\beta \beta_{0\nu,2B}$.
We may note that the effective couplings between neutrinos and the
majorons are easily estimated once the nuclear matrix elements are known.
Presently our main aim is to set up an upper limit on the neutrino mass from
the $0\nu$-decay analysis. Since a good understanding of the $0\nu$-decay
requires that we understand the $2\nu$-decay really well,
the $\beta \beta_{2\nu}$ decay rates have been reviewed as well.

The paper is organized as follows. In section \ref{sec2} we sketch the main
formalism and list various formulae needed to compute the half lives for the
$0^{+}\rightarrow 0^{+}$ transitions.
Section \ref{sec3} deals with the discussion of some important features of
nuclear matrix elements, the comparison of numerical results with the
experimental data, and the present limits on the neutrino mass. Finally, the
conclusions are drawn in section \ref{sec4}. The evaluation of radial
integrals, involved in the $0\nu$-decay, is shortly reviewed in appendix A.
\section{Formalism}
\label{sec2}
To apply the Horie-Sasaki method \cite{Hor61}, we write down
the Fourier-Bessel expansion of $0\nu$ transition amplitude as
\[
{\cal M}_{0{\nu}} =\frac{{\sf R}}{4\pi} \sum_{S \alpha J^{\pi} M}\hat{S}\int
d{\bf q}{\rm v}(q;{\omega}_{\alpha J^{\pi}})
\bra{0^+_f} e^{i{\bf qr}_1}\sigma^S_1{\rm t}_+ \ket{\alpha J^{\pi} M}
\otimes
\bra{\alpha J^{\pi}M}e^{-i{\bf qr}_2}\sigma^S_2 {\rm t}_+\ket{0^+_i},
\label{1}
\]
with
\be
{\rm v}(q;{\omega}_{\alpha J^{\pi}})= \frac{2}{\pi}
\frac{1}{q(q+{\omega}_{\alpha J^{\pi}})}
\label{1a}
\ee
being the neutrino potential, and
\[
{\omega}_{\alpha J^{\pi}}\cong {E}_{\alpha J^{\pi}}-(E_{i}+E_{f})/2.\label{1b}
\]
Here {\sf R} is the nuclear radius introduced to make ${\cal M}_{0{\nu}}$
dimensionless, the symbol $\otimes$ stands for the scalar product of
spin matrices as defined in ref. \cite{Boh69} (with $\sigma^{S}=1$ for $S=0$,
$\sigma^{S}=\sigma$ for $S=1$ and $\hat{S}=\sqrt{2S+1}$), the summation
goes over all virtual states $ \mid \alpha J^{\pi} M \rangle $, and
$ E_{i}$, ${E}_{\alpha J^{\pi}}$ and $E_{f}$ are the energies of the initial,
intermediate and final state, respectively.

After integrating over angles and summing up over M, we get
\begin{eqnarray*}
{\cal M}_{0{\nu}} & = & \sum_{LSJ^{\pi} } m_{0\nu} (L,S,J^{\pi})
\equiv 4 \pi {\sf R} \sum_{\alpha LSJ^{\pi} } (-)^{S}
\int dqq^{2}{\rm v}(q;{\omega}_{\alpha J^{\pi}})
\\
&\times&\Bra{0^{+}_{f}}{\cal O}_{+}(q{\bf r}_1;LSJ)\Ket{\alpha J^{\pi}}
\Bra{\alpha J^{\pi}}{\cal O}_{+}(q{\bf r}_2;LSJ)\Ket{0^{+}_{i}},
\label{3}
\end{eqnarray*}
where the quantities
\[
{\cal O}_{+} (q{\bf r};LSJ) =i^{L}\; j_{L} (qr) (Y_{L} \otimes
\sigma^{S})_{J}\; {\rm t}_{+},\label{4}
\]
are one-body operators.

Within the QRPA formulation presented in ref. \cite{Hir90a} the energies
$ {\omega}_{\alpha J^{\pi}} $ are  solutions  of  the  QRPA equation,
and the transition matrix elements are given by
\begin{eqnarray*}
\Bra{\alpha J^{\pi}}{\cal O}_{+} (q {\bf r};LSJ)\Ket{0^{+}_i}
&=&-\sum_{{\rm pn}} \Bra{{\rm p}} O (q {\bf r};LSJ)\Ket{{\rm n}}
\Lambda_{+} ({\rm pn};\alpha J^{\pi})
\\
\Bra{0^{+}_{f}}{\cal O}_{+} (q {\bf r};LSJ)\Ket{\alpha J^{\pi}}
&=&-\sum_{{\rm pn}} \Bra{{\rm p}} O (q{\bf r};LSJ)\Ket{{\rm n}}
\Lambda^{*}_{-} ({\rm pn};\alpha J^{\pi}),
\label{5}
\end{eqnarray*}
where the reduced {\rm pn} form factors are
\[
\sqrt{4 \pi} \Bra{{\rm p}} O (q {\bf r};LSJ)\Ket{{\rm n}}
= W^{LSJ}_{{\rm pn}} R^L_{{\rm pn}} (q) ,
\label{6}
\]
with the radial part
\[
R^L_{{\rm pn}}(q)=\int^{\infty}_{0}u_{\rm n}(r) u_{\rm p}(r)j_L(qr)r^{2}dr,
\label{7}
\]
{\it u}(r) being the single-particle radial wave functions, and the
angular part
\[
W^{LSJ}_{{\rm pn}} = i^{\ell_{{\rm n}}-\ell_{{\rm p}}+L}\sqrt{2}\, \hat{J}\,
\hat{S}
\,\hat{L} \,\hat{j}_{{\rm p}} \,\hat{j}_{{\rm n}}\, \hat{\ell}_{{\rm n}}\,
(\ell_{{\rm n}} 0 L 0\mid \ell_{{\rm p}} 0)
\left\{ \begin{array}{ccc}
\ell_{{\rm n}} & \frac{1}{2} & j_{{\rm n}} \\
L        &    S        &  J    \\
\ell_{{\rm p}} & \frac{1}{2} & j_{{\rm p}} \\
\end {array} \right\} .\label{8}
\]
The amplitudes $\Lambda_{\pm} ({\rm pn};\alpha J^{\pi})$  are defined as:
\begin{eqnarray*}
\L_{+}({\rm pn};\alpha J)=\sqrt{\rop \ron}
\left[\up \vn X_{{\rm pn};\alpha J}+\vbp \ubn Y_{{\rm pn};\alpha J}\right],
\\
\L_{-}({\rm pn};\alpha J)=\sqrt{\rop \ron}
\left[\vbp \ubn X_{{\rm pn};\alpha J}+\up \vn Y_{{\rm pn};\alpha J}\right],
\label{8a}
\end{eqnarray*}
where the unbarred (barred) quantities indicate that the quasiparticles
are defined with respect to the initial (final) nucleus;
$\rop^{-1}=\up^{2}+\vbp^{2}$, $\ron^{-1}=\ubn^{2}+\vn^{2}$,
and all the remaining notation has the standard meaning \cite{Hir90b,Krm92}.
As such, we obtain
\begin{eqnarray}
m_{0\nu} (J^{\pi})& = & \sum_{\alpha {\rm pnp}'{\rm n}'}
\Lambda_{+} ({\rm pn};\alpha J^{\pi}) \Lambda^{*}_{-} ({\rm p}'{\rm n}';\alpha
J^{\pi})
\nonumber\\
&\times& \sum_{LS}(-)^{S}  W^{LSJ}_{{\rm pn}}  W^{LSJ}_{{\rm p}'{\rm n}'}
{\cal R}^{L} ({\rm pn}{\rm p}'{\rm n}';{\omega}_{\alpha J^{\pi}}),
\label{9}
\end{eqnarray}
with
\be
{\cal R}^{L}({\rm pn}{\rm p}'{\rm n}';{\omega}_{\alpha J^{\pi}})=
{\sf R}\int^{\infty}_{0} dqq^{2} {\rm v}(q;{\omega}_{\alpha J^{\pi}})
R^L_{{\rm pn}}(q)R^L_{{\rm p}'{\rm n}'} (q).
\label{10}
\ee
The expression (\ref{9}), for the evaluation of the $0\nu$ moments within
the QRPA, is much simpler, regarding the angular momentum recoupling, than the
ones currently used in the literature \cite{Tom87,Mut89,Mut91,Tom91}.
The radial integrals (\ref{10}) are of the same type as the ones that appear
in the work of Horie and Sasaki \cite{Hor61}, so we use their method to
evaluate these, without getting involved with Moshinsky brackets.
To show the simplicity of this procedure, a  few results relevant to this
work are  reviewed in appendix A.

In the present work we consider both the GT and Fermi $2\nu$ transition matrix
elements with the corresponding  amplitudes written in the form:
\begin{eqnarray}
{\cal M}_{2{\nu}}& =& \sum_{J=0,1} m_{2\nu} (J^{+})
\nonumber\\
 & \equiv & \sum_{J=0,1}(-)^{J}  \sum_{\alpha {\rm pnp}'{\rm n}'}
\Lambda_{+} ({\rm pn};\alpha J) \Lambda^{*}_{-} ({\rm p}'{\rm n}';\alpha J)
W^{0JJ}_{{\rm pn}}W^{0JJ}_{{\rm p}'{\rm n}'}/{\omega}_{\alpha J^{+}}.
\label{11}
\end{eqnarray}
The corresponding total $\beta^{\mp}$ transition strengths are:
\begin{equation}
{\cal S}_{\pm}=\sum_{\alpha}
\left|\sum_{{\rm pn}}\Lambda_{\pm}({\rm pn};\alpha J) W^{0JJ}_{{\rm
pn}}\right|^2.
\label{11a}
\end{equation}
For the discussion of the numerical results the following
unperturbed moments are also needed:
\begin{eqnarray}
 m_{0\nu}^{0} (J^{\pi}) = \sum_{{\rm pn} L S}(-)^{S}
\Lambda_{+}^{0} ({\rm pn}) \Lambda_{-}^{0} ({\rm pn})
\left( W^{LSJ}_{{\rm pn}} \right)^{2}
{\cal R}^{L} ({\rm pn}{\rm pn};{\omega}^{0}_{{\rm pn}J^{\pi}}),\label{12}
\end{eqnarray}
and
\begin{eqnarray}
 m_{2\nu}^{0} (J^{+}) =(-)^J  \sum_{{\rm pn} }
\Lambda_{+}^{0} ({\rm pn}) \Lambda_{-}^{0} ({\rm pn})
\left(W^{0JJ}_{{\rm pn}} \right)^2
/{\omega}^{0}_{{\rm pn} J^{+}}, \label{13}
\end{eqnarray}
where ${\omega}^{0}_{{\rm pn}J^{\pi}}$ are the unperturbed pn-energies
for a given $J^{\pi}$, and
\[
\L_{+}^{0}({\rm pn})=\sqrt{\rop \ron}\,\up \vn\:,\:
\L_{-}^{0}({\rm pn})=\sqrt{\rop \ron}\,\ubn \vbp.
\label{13a}
\]

Finally, when the RHC mechanism is not considered,
the half lives for the $0^{+}\rightarrow 0^{+}$ transitions read
\[
T_{1/2}={\cal G}^{-1}({\cal M F})^{-2},
\]
where ${\cal G}$ is a kinematical factor \cite{Doi85,Doi88a}, ${\cal M}$ is
the nuclear matrix element, and the values of ${\cal F}$ are
\begin{eqnarray*}
&&{\cal F} =  \left\{
\begin{array}{cc}
1&\;\;\mbox{for}\;{\beta \beta}_{2\nu} \;,\\
\frac{<m_{\nu}>}{m_e}&\;\;\mbox{for}\;{\beta \beta}_{0\nu} \;.\\
\end{array}\right.
\end{eqnarray*}
\section {Numerical Results and Discussion}
\label{sec3}
\subsection{Unperturbed Matrix Elements}
\label{sec31}
As in our previous studies of the $\beta \beta $-decays \
\cite{Hir90,Hir90a,Hir90b,Krm90,Krm92}, the numerical calculations are
performed with a $\delta$-force (in units of $MeV fm^{3}$)
\[
V=-4\pi({\it v}_{s}P_{s}+{\it v}_{t}P_{t})\delta(r),
\]
with different strength constants ${\it v}_{s}$ and  ${\it v}_{t}$ for the
particle-hole, particle-particle and pairing channels.

For the nuclei $^{76}Ge$,  $^{82}Se$, $^{128}Te$, $^{130}Te$ and
$^{100}Mo$, we work in an eleven dimensional model space including all the
single particle orbitals of oscillator shells $3\hw$ and $4\hw$ plus the
$0h_{9/2}$ and $0h_{11/2}$ orbitals from the $5\hw$ oscillator shell. The
single particle energies (s.p.e.), as well as the parameters
${\it v}_{s}^{pair}({\rm p})$ and  ${\it v}_{s}^{pair}({\rm n})$, have been
fixed by
the procedure employed in ref.~\cite{Hir90a} (i.e., by fitting the experimental
pairing gaps  to a Wood-Saxon potential well). In fact, for the
nuclei $^{76}Ge$, $^{82}Se$, $^{128}Te$, and $^{130}Te$ we use here the same
parameterization as in ref.\ \cite{Hir90a}.

 For the nucleus $^{48}Ca$ the calculations have been done with three
different model spaces, namely those including all the orbitals in the
major shells: $2\hw$ and $3\hw$ ({\em space A}),  $0\hw$ to $3\hw$
({\em space B}), and $0\hw$  to $4\hw$ ({\em space C}).
Experimental single particle energies have been used for the orbitals,
$1p_{1/2}$, $0f_{5/2}$, $1p_{3/2}$, $0f_{7/2}$, $1s_{1/2}$ and
$0d_{3/2}$, here. For the remaining orbitals a single-particle
energy spacing of $\hw=41 A^{-1/3}$ MeV is assumed.

The unperturbed $2\nu$ and $0\nu$ moments, given by eqs. (\ref{12}) and
(\ref{13}) and displayed in table \ref{table2},  represent the upper limit to
the perturbed moments. We discuss these first, in order to  gain an insight
into both the magnitude of $\beta \beta$ moments and the role played by the
size of the configuration space. The $0\nu$ matrix elements have been
calculated by taking 5 MeV as the mean excitation energy ${\omega}_{\alpha
J^{\pi}}$ in the neutrino potential (\ref{1a}). It has been found that the
largest fraction of the $0\nu$ strengths for all the nuclei is concentrated
around this energy.

We may note  that, while the ${\it m}_{2\nu}^0(J^{\pi}) $ matrix elements are
practically unchanged as we go to larger configuration spaces, the ${\it
m}_{0\nu}^0(J^{\pi}) $ matrix elements, and in particular those with $J \geq
2$, vary significantly. The total moment $\M_{0\nu}^0 $, on the other hand,
shows only a small variation, due to the dominance of the
${\it m}_{0\nu}^0(0^{+})$ and ${\it m}_{0\nu}^0(1^{+})$ in comparison with
the moments ${\it m}_{0\nu}^0( J \geq 2)$. This result is not surprising as,
from the study of the charge-exchange resonances, we already know that for
$^{48}Ca$ the spaces A and B are complete spaces only for the Fermi and GT
transitions. For the first-forbidden resonances
$(L=1,J^{\pi}=0^{-},1^{-},2^{-})$, a calculation that does not include the
$4\hw$ shell as well, would  be totally unacceptable. In the same way for the
evaluation of the second-forbidden resonances with $L=2$ and $J^{\pi}=1^{+},
2^{+},3^{+}$ one should include the orbitals of the $5\hw$ oscillator shell
(otherwise the corresponding charge exchange sum rule will not be satisfied),
and so on.  Fortunately, the problem with the size of the configuration space
is not so serious for the $\beta \beta$-decay.  The reason is that
the transition amplitude for this process is proportional to the factor
$\up\vn\ubn\vbp$ that rapidly decreases as we move away from the valence
orbitals. The same argument is not valid for charge-exchange transitions whose
amplitudes carry the factor $\up\vn$ or the factor $\ubn\vbp$.
\subsection{Perturbed Matrix Elements}
\label{sec32}

The particle-hole channel parameter values, ${\it v}_s^{ph}=27$ and ${\it
v}_t^{ph}=64$ for $^{48}Ca$ and ${\it v}_s^{ph}=55$ and ${\it v}_t^{ph}=92$ for
the remaining six nuclei, have been taken from a study of energy systematics of
the GT resonances (\cite{Nak82,Cas87}. For further discussion it is convenient
to introduce the parameters $s$ and $t$, defined as the ratios between the
$T=1$, $S=0$ and $T=0$, $S=1$ coupling constants in the PP channels and the
pair
   ing
force  constants, i.e.,
\[
s=\frac{2{\it v}_s^{pp}}{{\it v}_s^{pair}({\rm p})+{\it v}_s^{pair}({\rm
n})}\quad;\quad
t=\frac{2{\it v}_t^{pp}}{{\it v}_s^{pair}({\rm p})+{\it v}_s^{pair}({\rm
n})}.\\
\]

For a value of $s \cong 1$ the isospin symmetry is
restored within the QRPA, leading to a concentration of $\S_{+}(0^{+})$
strength in a single state and resulting in
\begin{equation}
\S_{-}(0^{+})\cong0\;;\; {\it m}_{2\nu}(0^{+})\cong0,\;\;
{\rm and}\;\; {\it m}_{0\nu}(0^{+})\cong0.\label{21}
\end{equation}

{}From  fig.~\ref{figure1} it is evident that  the conditions
(\ref{21}) are fulfilled reasonably well for all the nuclei discussed here.
As such we fix the spin-singlet PP strength at the value $s=1$,
obtaining ${\it m}_{2\nu}(0^{+}) \cong 0$ and $\M_{2\nu} \cong {\it
m}_{2\nu}(1^{+})$.  It should be noted that in most of the QRPA calculations
performed so far, the moment ${\it m}_{2\nu}(0^{+})$ has been simply ignored by
invoking the isospin symmetry. Simultaneously, however, sizable values
for the moment ${\it m}_{0\nu}(0^{+})$ have been reported. Apparently
such calculations are inconsistent since the restoration of the
isospin symmetry within the QRPA takes place at the level of the residual
interaction and not at the level of the one body operator ${\rm t}_{+}$.

Before looking at the experimental data  we consider the question of
validity of the BCS approximation for far off orbitals as the configuration
space is extended. With this in mind, three different calculations  have been
performed for $^{48}Ca$, namely\\
{\em Calculation I}: Both the BCS and the QRPA equations have been solved
within the single-particle space A.\\
{\em Calculation II}: The configuration space for solving QRPA equations
extends over the major shells  $0\hw-4\hw$ (space C), but the gap equations
have been solved within the space B only. The $4\hw$ shell is assumed to be
totally empty.\\
{\em Calculation III}: The full space C has been used at all steps of
the calculation.\\
In each case, the pairing interaction strengths have been determined by fitting
the experimental odd-even mass differences for the relevant nuclei.
The calculated $\S_{-}(1^{+})$, $-{\it m}_{2\nu}(1^+)$,
${-\it m}_{0\nu}(1^{+})$ and $-\M_{0\nu}$, as a function of parameter $t$,
are shown in fig. \ref{figure2}. The curves have been drawn up to the first
pole of ${\it m}_{2\nu}(1^+)$, where the energy of the lowest virtual
$1^{+}$ state becomes equal to the energy of the initial or final state
and the QRPA breaks down. The magnitudes of all the quantities mentioned above
depend to some extent on the size of the configuration space and the way in
which the BCS equations have been solved. Yet, the general trend is the same
in the three spaces. As a matter of fact, if one fixes the value of the
parameter $t$ at the lowest value of $\S_{-}(1^{+})$, we get practically the
same results from all three calculations; that is:
${\it m}_{2\nu}(1^+)\cong0.09$ $MeV^{-1}$, ${\it m}_{0\nu}(1^{+})\cong0.9$
$MeV^{-1}$ and $\M_{0\nu}\cong-0.45$ $MeV^{-1}$.

We may generalize the conclusions drawn for the case of  $^{48}Ca$.
That is, an enlargement of the configuration space, beyond two major oscillator
shells, is not expected to modify in essence the model predictions. As such the
11 single-particle space described earlier should be a rather complete space
for the description of the $\beta \beta$ processes in the remaining five
nuclei.
Fig. \ref{figure3} shows the relevant results for the nuclei $^{76}Ge $,
$^{82}Se$, $^{100}Mo$, $^{128}Te$ and $^{130}Te$.  From an inspection of figs.
\ref{figure2} and \ref{figure3}  the following common points may be made:\\
1) The minimum of $\S_{-}(1^{+})$ always occurs at a value of $t=t_{sym}$ that
lies between the zero and the pole of ${\it m}_{2\nu}(1^+)$.\\
2) As pointed out earlier in a similar context
(\cite{Vog86,Civ87,Tom87,Eng88,Mut88,Mut89}), in the vicinity of $t_{sym}$,
the moment ${\it m}_{0\nu}(1^{+})$ is very sensitive to small variations of
the parameter {\it t}, while the total moment $\M_{0\nu}$ is only moderately
influenced by the same variation. The only exception is the nucleus $^{48}Ca$.
Here, due to a very pronounced dominance of the ${\it m}_{0\nu}(1^{+})$ moment
over the moments with higher multipolarities, the total $0\nu $ matrix element
passes through zero at a value of {\it t} that is very close to $t_{sym}$.
\subsection{Comparison with Experimental Data}
\label{sec33}
The measured $2\nu$ and $0\nu$  half-lives are listed in table \ref{table3}
along with the values of kinematical factors $G^{2 \nu}$ and $G^{0 \nu}$,
appropriately renormalized for an effective axial-vector coupling constant
$g_{A}=-g_{V}$, and the resulting observables $|\M_{2\nu}|$ and
$|{\cal M}_{0\nu}\frac{<m_{\nu}>}{m_{e}}|$.

The results of the calculations for the matrix elements $\M_{2\nu}$ and
$\M_{0\nu}$, as well as for the predicted $2\nu$ half-lives and the
neutrino masses, are shown in table \ref{table4}.
{}From the upper panel of this table it is seen that when $t= t_{sym}$ the
calculated ${2\nu}$  matrix element for $^{48}Ca$ does not contradict the
experimental limit and that the measurement of the moment $\M_{2\nu}$ in
$^{82}Se$ is well accounted for by the theory.
But, the calculated matrix elements $\M_{2\nu}$ turn out to be too small for
$^{76}Ge$ and $^{100}Mo$ and too large for $^{128}Te$ and $^{130}Te$
(in both the cases by a factor of $\approx 3$).
It is worth noting that the measured values of $|\M_{2\nu}|$ for $^{76}Ge$
and $^{100}Mo$ nuclei are not very different from the calculated ones even
when no PP interaction is included (see fig.~\ref{figure3}).
{}From  the same figure, we may notice that  the minima of $\S_{-}(1^{+})$ are
not so well defined and it is precisely near these minima that
the calculated values of $\M_{2\nu}$ vary rather abruptly.
Evidently, as such, the experimental data for the $2\nu$ half-lives can
be reproduced in all the six nuclei with a value of $t$ very close to
$t= t_{sym}$.
Finally, the calculated $\M_{0\nu}$ moments (and estimated neutrino mass
limits) for $t$-values that reproduce the measured matrix elements
${\cal M}_{2\nu}$, when these are assumed to be positive ($t=t_{\uparrow}$)
and negative ($t=t_{\downarrow}$), are listed in middle and lower panels of
table \ref{table4}, respectively.
One notices that in all the cases $t_{sym}\cong t_{\uparrow}$, and $t_{sym}$
and $t_{\uparrow}$ lead to practically the same upper limits for effective
neutrino mass $<\!m_{\nu}\!>\cong 1\,eV$.
\section {Conclusions}
\label{sec4}
We have calculated the $2\nu$ and $0\nu$ double beta observables for the nuclei
$^{48}Ca$, $^{76}Ge$, $^{82}Se$, $^{100}Mo$, $^{128}Te$ and  $^{130}Te$ in the
framework of the QRPA model satisfying the constraints imposed on the
particle-particle coupling strengths by the isospin and SU(4) symmetries,
i.e., $s=1$ and $t\cong t_{sym}$.
With the parameter $t$  equal to  $t_{sym}$ the calculations are consistent
with the experimental limit for the $2\nu$ lifetime of $^{48}Ca$ and they
correctly reproduce the $2\nu$ lifetime of $^{82}Se$.
For the remaining nuclei the $2\nu$ half lives are either overestimated
(for $^{76}Ge$ and $^{100}Mo$) or underestimated (for $^{128}Te$ and
$^{130}Te$)
by an order of magnitude.
This does not imply that one has to forgo the idea of reconstructing the SU(4)
symmetry by the residual interaction.  We believe that the restoration of
both the isospin and SU(4)  symmetries is a genuinely useful feature of
QRPA and undoubtedly plays an important role in the intricate physics
involved in $\beta \beta$ processes.  One also should bear in mind that:
i) with a comparatively small variation ( $<10\%$) of $t$ with respect to
$t_{sym}$, i.e., with $t=t_{\uparrow}$ it is possible to account for the
$T_{1/2}^{2\nu}$ in all the cases, and
ii) the minimum value of the GT ${\cal S}_{-}(1^{+})$ strength critically
depends on the spin-orbit splitting over which we still do not have a
complete control.  In the same context it should be interesting to study
the role of forbidden virtual states, particularly those with
$J^{\pi}=1^{-}$ and $2^{-}$ (see table~\ref{table2}), in building up the
total ${\cal M}_{2\nu}$ moments \cite{Wil88}.

For a long time we have been worrying about the  completeness of the virtual
states in the evaluation of the $0\nu$ moments, or in other words: how the
$0\nu$ moments depend on the size of the configuration space? The results
displayed in table~\ref{table2} and fig.~\ref{figure2} seem to suggest that
the enlargement of the space beyond two oscillator shells does not have a
significant  effect either on the  GT strength $\S_{-}(1^{+})$ or on the
$\beta\beta$ moments  ${\it m}_{2\nu}(1^+)$, ${\it m}_{0\nu}(1^{+})$ and
$\M_{0\nu}$. In the first case the configuration space turns out to be
sufficiently complete while in the second case  the pairing factor dependence
of the $\beta\beta$ transition amplitudes inhibits the far off orbitals to
contribute. We feel, however, that this question is not yet resolved
unambiguously.

A final and rather general comment is in order.
Besides the issue of the procedure adopted for fixing the particle-particle
strength
parameter, there are some additional problems within the QRPA calculations of
the matrix element ${\cal M}_{2\nu}$, as yet not fully understood.
They are related with the type of force, choice of the single
particle spectra, treatment of the difference between the initial and final
nuclei, etc. All these things are to some extent uncertain and therefore it
is open to question whether it is possible, at the present time, to obtain
a more reliable theoretical estimate for the $2\nu$ half lives that the one
reported here.
Similar remarks stand for the ${\cal M}_{0\nu}$ moments and hence for
the neutrino mass limits. The difference in a factor of about $2-3$ between
both:
i) the results obtained by the Pasadena group and the groups of T\"{u}bingen
and Heidelberg for $^{76}Ge$ and $^{82}Se$ nuclei, and ii) the previous and
present calculations  for $^{100}Mo$, $^{128}Te$ and $^{130}Te$ nuclei,
is just a reflection of the unavoidable uncertainty of the QRPA calculations,
and it is difficult to assess which one is "better" and which is "worse".
\newpage
\appendix
\section{Radial Matrix Elements for the $0\nu$$ \beta \beta$-decay}

     Within the  Horie-Sasaki formalism \cite{Hor61}  the radial
matrix elements (\ref{10}) for the  harmonic oscillator wave
functions read as
\begin{eqnarray*}
{\cal R}^{L} ({\rm pnp}'{\rm n}';{\omega}_{\alpha J^{\pi}})
&=&{\sf R}\left[M({\rm p},{\rm n}) M({\rm p}',\rm {n}')\right]^{-\frac{1}{2}}
\\
& \times & \sum_{mm'} a_{m}({\rm p},{\rm n}) a_{m'}({\rm p}',{\rm n}')
f^{L}(m,m';{\omega}_{\alpha J^{\pi}}),
\end{eqnarray*}
with
\[
M(n\ell,n'\ell')= 2^{n+n'} n! n'!
(2\ell+2n+1)!! (2\ell'+2n'+1)!!,
\]
\[
a_{\ell+\ell'+2s}(n\ell, n'\ell')=\sum_{(k+k'=s)}
\beg{n}{k} \beg{ n'}{k'}
\frac{(2 \ell+2n+1)!!}{(2\ell+2k+1)!!}
\frac{(2 \ell'+2n'+1)!!}{(2 \ell'+2k'+1)!!},
\]
\[
f^L(m,m';{\omega}_{\alpha J^{\pi}})=\sum_{\mu} a_{2 \mu}
\left(\frac{m-L}{2} L, \frac{m'-L}{2} L \right)
{\cal J}_{\mu}({\omega}_{\alpha J^{\pi}}),
\]
and
\[
{\cal J}_{\mu}({\omega}_{\alpha J^{\pi}})= (2\nu)^{-\mu}\int^{\infty}_{0}
dqq^{2\mu+2}exp(-q^{2}/{2\nu}){\rm v}(q;{\omega}_{\alpha J^{\pi}}),
\]
where $\nu=M\omega/\hbar$ is the oscillator parameter. For
${\rm v}(q;{\omega}_{\alpha J^{\pi}})$  given  by  eq. (\ref{1a}),
Tomoda et al.\ \cite{Tom86} have obtained the following recurrence relation
for the momentum space integrals
\[
{\cal J}_{\mu}({\rm u})=\sqrt{\frac{2\nu}{\pi}}\frac{(2\mu-1)!!}{2^{\mu}}
-\frac{\sqrt{2\nu}}{\pi} {\rm u}(\mu-1)!+{\rm u}^2{\cal J}_{\mu-1}({\rm u}),
\]
\[
{\cal J}_{0}({\rm u})=\sqrt{\frac{2\nu}{\pi}}
-\frac{2\sqrt{2\nu}}{\pi}{\rm u}\Phi({\rm u}),
\]
where ${\rm u}={\omega}_{\alpha J^{\pi}}/\sqrt{2\nu}$ and
\[
\Phi({\rm u})=\int^{\infty}_{0}\frac{exp(-t^{2})}{t+{\rm u}}dt=exp(-{\rm u}^2)
\left[\sqrt{\pi}\int^{\infty}_{0}exp(t^{2})dt-\frac{1}{2}Ei({\rm u}^2)\right].
\]

When finite nucleon size (FNS) effect and the short range (SR)
two-nucleon correlations are included, the potential
${\rm v}(q;{\omega}_{\alpha J^{\pi}}) $ takes the form
\[
{\rm v}_{FNS+SR}(q;{\omega}_{\alpha J^{\pi}})=
{\rm v}_{FNS}(q;{\omega}_{\alpha J^{\pi}})
-\Delta {\rm v}(q) +\Delta'{\rm v}(q),
\]
with
\[
{\rm v}_{FNS}(q;{\omega}_{\alpha J^{\pi}})=
{\rm v}(q;{\omega}_{\alpha J^{\pi}})
\left(\frac{\Lambda^{2}}{\Lambda^{2}+q^{2}}\right)^{4}\:,\:
\Delta {\rm v}(q)=\frac{2\pi}{qq_{c}}
ln \left| {\frac{q+q_{c}}{q-q_{c}}}\right|,
\]
\[
\Delta {\rm v}'(q)=\frac{\pi}{qq_{c}}\left[\sum_{n=1}^{3}\frac{1}{n}
\left(x_{-}^{n}-x_{+}^{n}\right)+ln\left(\frac{x_{-}}{x_{+}}\right)\right]\:;\:
x_{\pm}=\frac{\Lambda^{2}}{\Lambda^{2}+(q \pm q_{c})^{2}},
\]
where $ \Lambda=850 MeV $ is the cutoff for the dipole form factor and
$q=3.93\,fm^{-1}$ is roughly the Compton wavelength of the $\omega$-meson.
The corresponding integrals $ {\cal J}_{\mu}({\omega}_{\alpha J^{\pi}}) $
have to be evaluated numerically.

%
\newpage

\bigskip

\begin{figure}[t]
\begin{center} { \large Figure Captions} \end{center}
\caption{Fermi observables ${\cal S}_{-}(J^{\pi}=0^{+})$,
${\it m}_{2\nu}(J^{\pi}=0^{+})$ (in units of $[MeV]^{-1}$)
and ${\it m}_{0\nu}(J^{\pi}=0^{+})$ for the nuclei $^{48}Ca$, $^{76}Ge $,
$^{82}Se$, $^{100}Mo$,  $^{128}Te$ and  $^{130}Te$, as a function of
particle-particle $S=0$, $T=1$ coupling constant $s$.}
\label{figure1}

\bigskip

\caption{Gamow-Teller observables ${\cal S}_{-}(J^{\pi}=1^{+})$,
${\it m}_{2\nu}(J^{\pi}=1^{+})$ (in units of $[MeV]^{-1}$),
 ${\it m}_{0\nu}(J^{\pi}=1^{+})$ and the total ${\cal M}_{0\nu}$ moment
 for the nucleus $^{48}Ca$, as a function of the particle-particle $S=1$,
$T=0$ coupling constant $t$, within the single-particle spaces A, B and C.}
\label{figure2}

\bigskip

\caption{Gamow-Teller observables ${\cal S}_{-}(J^{\pi}=1^{+})$,
${\it m}_{2\nu}(J^{\pi}=1^{+})$ (in units of $[MeV]^{-1}$),
${\it m}_{0\nu}(J^{\pi}=1^{+})$ and the total ${\cal M}_{0\nu}$ moment for
the nuclei $^{76}Ge $, $^{82}Se$, $^{100}Mo$, $^{128}Te$ and $^{130}Te$,
as a function of the particle-particle $S=1$, $T=0$ coupling constant $t$.}
\label{figure3}
\end{figure}

\bigskip

\mbox{}
\newpage


\begin{table}[t]

\begin{center} {\large Tables }

\caption {Upper bounds on the effective neutrino mass $<m_{\nu}>$ (in eV)
obtained from the QRPA calculations of the nuclear matrix elements. For the
sake of comparison, in all the cases the effective axial vector coupling
constant $g_{A}^{eff}=-g_{V}$ has been employed. This means that the results
for $<m_{\nu}>$ from refs.\ \protect \cite{Tom87,Mut89,Mut91,Suh91a} have been
properly
renormalized by the factor $(1.25)^{2}$. The experimental data for the
half-lives which have been used are indicated in the second row.
The two values of the Pasadena group correspond to their PP strengths:
(a)  ${\alpha}_{1}^{'}=-390$  and (b) ${\alpha}_{1}^{'}=-432$, in units
of MeV $fm^3$.}
\label{table1}

\bigskip

\begin{tabular}{lccccccc}
\hline
\hline

&{$^{48}Ca$}&{$^{76}Ge$}&{$^{82}Se$}&{$^{100}Mo$}&{$^{128}Te$}&{$^{130}Te$}\\
Exp.&ref. \cite{You91} &ref. \cite{Bal92} &ref. \cite{Ell92}
&ref. \cite{Eji91} &ref. \cite{Ber92} &ref. \cite{Ale92}\\
\hline

ref.\ \cite{Tom87}                     &&$2.3$&$8.2$&&$2.4$&$24$\\
ref.\ \cite{Eng88} (a)           &&$4.4$&$20$&$20$&$1.8$&$22$\\
ref.\ \cite{Eng88} (b)                &&$10$&$41$&$$&$2.4$&$29$\\
ref.\ \cite{Mut89}                       &&2.0&7.4&26&1.5&21\\
ref.\ \cite{Mut91}                       &$22$&&&&&\\
ref.\ \cite{Suh91a}                   &&$3.1$&$12$&&$3.8$&$31$\\
\hline
\hline

\end{tabular}
\end{center}
\end{table}

\newpage

\begin{table}[t]

\begin{center}

\caption{Unperturbed ${\it m}_{2\nu}(J^{\pi})$ and ${\it m}_{0\nu}(J^{\pi})$
moments in units of $[MeV]^{-1}$. As explained in the text, three different
single-particle spaces have been used for the nucleus $^{48}Ca$.}
\label{table2}

\bigskip

\begin{tabular}{c|cccrrrrrr}\hline
\hline

&\multicolumn{3}{c}{$^{48}Ca$}& {$^{76}Ge$}& {$^{82}Se$}& {$^{90}Mo$}&
{$^{128}Te$}& {$^{130}Te$}\\
&A&B&C&&&&&\\
\hline
${-\it m}_{2\nu}^0 (J^{\pi})$ &&&&&&&&\\
$0^{+}$    &       0.249 &0.255 &0.253 &0.423 &0.491 &0.256 &0.678 &0.615 \\
$1^{+}$    &       0.406 &0.417 &0.414 &0.923 &0.966 &1.513 &1.602 &1.434 \\
\hline
${-\it m}_{0\nu}^0 (J^{\pi})$ &&&&&&&&\\
$0^{+}$    &       0.928 &0.953 &0.990 &2.258 &2.424 &1.857 &2.671 &2.472 \\
$1^{+}$    &       2.072 &2.158 &2.268 &6.168 &6.259 &7.204 &7.310 &6.751 \\
$2^{+}$    &       0.316 &0.384 &0.426 &1.285 &1.283 &1.470 &1.483 &1.367 \\
$3^{+}$    &       0.321 &0.394 &0.438 &1.112 &1.179 &1.421 &1.260 &1.164 \\
$4^{+}$    &       0.109 &0.131 &0.151 &0.507 &0.525 &0.708 &0.636 &0.587 \\
$5^{+}$    &       0.134 &0.154 &0.170 &0.407 &0.459 &0.636 &0.497 &0.464 \\
$6^{+}$    &       0.037 &0.037 &0.044 &0.196 &0.214 &0.356 &0.253 &0.236 \\
$7^{+}$    &       0.075 &0.075 &0.079 &0.145 &0.187 &0.275 &0.212 &0.200 \\
$8^{+}$    &       0.000 &0.000 &0.001 &0.055 &0.070 &0.160 &0.096 &0.092 \\
$9^{+}$    &       0.000 &0.000 &0.000 &0.056 &0.083 &0.065 &0.087 &0.084 \\
$10^{+}$   &       0.000 &0.000 &0.000 &0.002 &0.002 &0.006 &0.029 &0.028 \\
$0^{-}$    &       0.027 &0.040 &0.071 &0.177 &0.194 &0.337 &0.130 &0.126 \\
$1^{-}$    &       0.307 &0.378 &0.564 &1.604 &1.749 &2.655 &1.367 &1.315 \\
$2^{-}$    &       0.209 &0.250 &0.337 &1.161 &1.266 &1.628 &1.128 &1.078 \\
$3^{-}$    &       0.158 &0.183 &0.244 &0.882 &0.954 &1.124 &0.929 &0.881 \\
$4^{-}$    &       0.061 &0.061 &0.093 &0.517 &0.559 &0.735 &0.559 &0.531 \\
$5^{-}$    &       0.062 &0.063 &0.087 &0.417 &0.453 &0.456 &0.479 &0.453 \\
$6^{-}$    &       0.000 &0.000 &0.010 &0.157 &0.175 &0.276 &0.249 &0.237 \\
$7^{-}$    &       0.000 &0.000 &0.009 &0.181 &0.197 &0.171 &0.258 &0.245 \\
$8^{-}$    &       0.000 &0.000 &0.000 &0.014 &0.019 &0.051 &0.084 &0.082 \\
$9^{-}$    &       0.000 &0.000 &0.000 &0.015 &0.018 &0.048 &0.128 &0.122 \\
\hline
${-\cal M}_{0\nu}^{0}$& 4.816 &5.261
&5.983&17.317&18.268&21.641&19.844&18.516\\
\hline
\hline
\end{tabular}
\end{center}
\end{table}

\newpage

\begin{table}[t]

\begin{center}

\caption {Values of the measured half-lives ${T_{1/2}}$, kinematical factors G
and the experimental  ${\cal M}F$-quantities for the ${2\nu}$ and ${0\nu}$
${\beta\beta}$ decays. The G factors are those from ref. \protect
\cite{Doi88a},
but renormalized  for $g_{A}=-g_{V}$.}
\label{table3}

\bigskip

\begin{tabular}{cccccccc}
\hline
\hline

Observable&{$^{48}Ca$}& {$^{76}Ge$}& {$^{82}Se$}& {$^{100}Mo$}&
{$^{128}Te$}& {$^{130}Te$}\\
\hline

${T_{1/2}^{2\nu} [yr \; 10^{20}]}$ & $\!>0.36\,^{a}\!$
                                       & $\!9.2_{-0.4}^{+0.7}   \,^{b}\!$
                        & $\!1.08_{-0.06}^{+0.26}\,^{c}\!$
                        & $\!0.115_{-0.020}^{+0.030}\,^{d}\!$
                        & $\!(7.7\pm0.4)             \; 10^{4}\,^{e}\!$
                        & $\!27\pm1             \,^{e}\!$\\
$\!G^{2\nu}[yr(MeV)^{2}]^{-1}\!$ &$\! 0.423\; 10^{-17}\!$
                                 &$\! 0.139\; 10^{-19}\!$
                                 &$\! 0.464\; 10^{-18}\!$
                                 &$\! 0.101\; 10^{-17}\!$
                                 &$\! 0.911\; 10^{-22}\!$
                                 &$\! 0.512\; 10^{-18}\!$\\
$|{\cal M}_{2\nu}|[MeV]^{-1}$      &  $   <0.081              $
                             &  $   0.280_{-0.010}^{+0.006}$
                             &  $   0.141_{-0.014}^{+0.004}$
                             &  $   0.294_{-0.033}^{+0.029}$
                             &  $   0.038_{-0.01}^{+0.01} $
                             &  $   0.027_{-0.01}^{+0.01} $\\
${T_{1/2}^{0\nu} [yr\; 10^{21}]}$ & $\!>9.5 \, ^{f}\!$
                                      & $\!>1400\, ^{g}\!$
                                      & $\!>27 \, ^{c}\!$
                                      & $\!>4.7 \, ^{d}\!$
                                      & $\!>7700\, ^{e}\!$
                                      & $\!>2.5 \, ^{h}\!$\\
$G^{0\nu}[yr]^{-1}$          &  $\!   0.260\; 10^{-13}\!$
                             &  $\!   0.261\; 10^{-14}\!$
                             &  $\!   0.114\; 10^{-13}\!$
                             &  $\!   0.187\; 10^{-13}\!$
                             &  $\!   0.746\; 10^{-15}\!$
                             &  $\!   0.181\; 10^{-13}\!$\\
$|{\cal M}_{0\nu}\frac{<m_{\nu}>}{m_{e}}|\;10^4$ & $ < 0.64$
                                                 & $ < 0.17$
                                                 & $ < 0.57$
                                                 & $ < 1.07$
                                                 & $ < 0.13$
                                                 & $ < 1.5 $\\
\hline
\hline
\end{tabular}
\end{center}
$^{a})$ (laboratory data) ref.\ \cite{Bar70}\\
$^{b})$ (laboratory data) ref.\ \cite{Avi91}\\
$^{c})$ (laboratory data) ref.\ \cite{Ell92}\\
 $^{d})$ (laboratory data) ref.\ \cite{Eji91}\\
 $^{e})$ (geochemical data) ref.\ \cite{Ber92}\\
 $^{f})$ (laboratory data) ref.\ \cite{You91}\\
 $^{g})$ (laboratory data) ref.\ \cite{Bal92}\\
 $^{h})$ (laboratory data) ref.\ \cite{Ale92}
\end{table}
\newpage

\begin{table}[t]

\begin{center}
\caption
{Calculated ${2\nu}$ and ${0\nu}$ moments and the corresponding upper limits
for the effective neutrino mass $<m_{\nu}>$.
Here $s=1$ and three different sets of parameter $t$ have been considered,
namely, $t=t_{sym}$, $t=t_{\uparrow}$ and $t=t_{\downarrow}$. The last two
reproduce the measured $2\nu$ matrix element (without taking error bars into
consideration) when these are assumed to be positive and negative,
respectively.
For $t=t_{sym}$, the calculated $T_{1/2}^{2\nu}$ values are also shown.}
\label{table4}

\bigskip

\begin{tabular}{lccccccc}\hline
\hline

&{$^{48}Ca$}& {$^{76}Ge$}& {$^{82}Se$}& {$^{100}Mo$}&
{$^{128}Te$}& {$^{130}Te$}\\
\hline

$t_{sym}$                     &1.50&1.25&1.30&1.50&1.40&1.40\\
$\left({\cal M}_{2\nu}\right)_{sym}[MeV]^{-1}$
                              &0.091&0.100&0.121&0.102&0.118&0.096\\
$\left({\cal M}_{0\nu}\right)_{sym}$
                              &$-$0.46&$-$5.7&$-$5.6&$-$6.2&$-$7.0&$-$6.6\\
$\left<m_{\nu}\right>_{sym}[eV]$
                              &71&1.5&5.3&8.8&1.0&12\\
$\left(T_{1/2}^{2\nu}\right)_{sym} [yr\; 10^{20}]$
                              &0.28&71&1.5&0.95&$7.9 \times 10^{3} $&2.1\\
\hline

$t_{\uparrow}$                     &1.48&1.35&1.39&1.53&1.35&1.30\\
$\left({\cal M}_{0\nu}\right)_{\uparrow}$
                              &$-$0.48&$-$4.5&$-$5.3&$-$6.9&$-$7.3&$-$7.0\\
$\left<m_{\nu}\right>_{\uparrow}[eV]$
                               &68&1.9&7.1&10&1.0&11\\
\hline

$t_{\downarrow}$                     &1.20&0.30&1.00&1.30&1.22&1.20\\
$\left({\cal M}_{0\nu}\right)_{\downarrow}$
                     &$-$1.41&$-$10.2&$-$7.7&$-$8.3&$-$8.1&$-$7.5\\
$\left<m_{\nu}\right>_{\downarrow}[eV]$
                              &23&0.87&4.2&6.6&0.86&10\\
\hline
\hline
\end{tabular}
\end{center}
\end{table}
\end{document}